\documentclass[aps,prl,reprint,floatfix,amsmath,amssymb,longbibliography]{revtex4-2}
\usepackage{graphicx,microtype}
\usepackage[hidelinks]{hyperref}
\hypersetup{pdfauthor={Fu-Guo Yang},pdftitle={Unified Dynamics of Black-Hole Interiors: From Geometric to Scalar-Kinetic Dominance}}
\begin{document}
\title{\resizebox{\textwidth}{!}{Unified Dynamics of Black-Hole Interiors: From Geometric to Scalar-Kinetic Dominance}}
\author{Fu-Guo Yang}
\email{yang-fuguo@outlook.com}
\affiliation{International Centre for Theoretical Physics Asia-Pacific, University of Chinese Academy of Sciences, Beijing 100190, China}
\begin{abstract}
We identify a unified dynamical picture connecting small-amplitude scalar input at the event horizon to scalar--Kasner interiors. Within boundary-supported stationary families, area contraction drives the transition from geometric to scalar kinetic dominance. For the Schwarzschild-like branch, contraction suppresses the relative curvature contribution, so the scalar kinetic term becomes dominant while the gradient remains small. For the Kerr--Newman-like branch, the scalar gradient is strongly amplified as area contraction carries the solution toward the would-be Cauchy-horizon zero, followed by a narrow transition to scalar kinetic dominance. Cancellation between the kinetic and geometric terms produces a positive minimum of a metric diagnostic, explaining inner-horizon loss dynamically. Fully coupled numerical solutions for the Schwarzschild-like branch and four-dimensional stationary axisymmetric Kerr--Newman interiors, including the Reissner--Nordstr\"om and Kerr limits, support this picture. Area contraction provides the common physical mechanism, while the would-be Cauchy-horizon zero determines the strength of gradient amplification and the localization of the transition.

\end{abstract}
\maketitle
\flushbottom
\setlength{\parskip}{0pt}
\setlength{\abovedisplayskip}{6pt plus 3pt}
\setlength{\belowdisplayskip}{6pt plus 3pt}
\setlength{\abovedisplayshortskip}{6pt plus 3pt}
\setlength{\belowdisplayshortskip}{6pt plus 3pt}
\setlength{\textfloatsep}{10pt}
\setlength{\floatsep}{6pt}
\setlength{\dbltextfloatsep}{10pt}
\setlength{\dblfloatsep}{6pt}
\setlength{\jot}{3pt}
\setlength{\abovecaptionskip}{4pt}
\setlength{\belowcaptionskip}{0pt}

How can small-amplitude scalar input at the event horizon reshape a black-hole interior? Cauchy-horizon instability has been studied through mass inflation~\cite{PI,Ori,Dafermos,HA} and rotating vacuum interiors~\cite{DL,Gurriaran}. Scalar backreaction can also increase the interior mass and alter the Kasner geometry without a Cauchy horizon~\cite{GuoJoshi,Frenkel}. Charged and rotating systems exhibit inner-horizon loss and Kasner behavior, in which spatial lengths evolve as powers of the remaining proper time~\cite{CLY,ALY,DPEMH,DPEMGB,Hartnoll,DHS,li,LSY,Gao,BHR,GLL}.

Stationary solutions with scalar hair provide a controlled setting for following the nonlinear interior from the event horizon toward the singularity. Boundary scalar sources deform Schwarzschild--AdS interiors into Kasner regimes~\cite{Frenkel}. Dynamical analyses of charged spherical interiors connect small horizon scalar data to nonlinear collapse and a spacelike Kasner singularity~\cite{vdm}. Asymptotically flat charged hairy black holes also exhibit scalar oscillations and Kasner transitions~\cite{DHS}. Rotating studies include fully nonlinear synchronized-hair interiors in five dimensions~\cite{BHR} and stationary scalar-haired solutions in three-dimensional anti-de Sitter space~\cite{GLL}. A final Kasner regime also arises in analyses allowing general anisotropy and spatial dependence~\cite{Henneaux}.

What common dynamics connects small scalar input to these interior changes? As the cross-sectional area of the ingoing null geodesic congruence shrinks, the scalar kinetic term can overtake the other terms in the radial equations. The kinetic term can become dominant before its accumulated effect produces an order-one metric change. We determine these two radial scales and, when a Cauchy horizon is present in the background, identify the balance that replaces the would-be horizon zero with a positive minimum of the metric diagnostic.

\begin{figure*}[t]
\includegraphics[width=\textwidth]{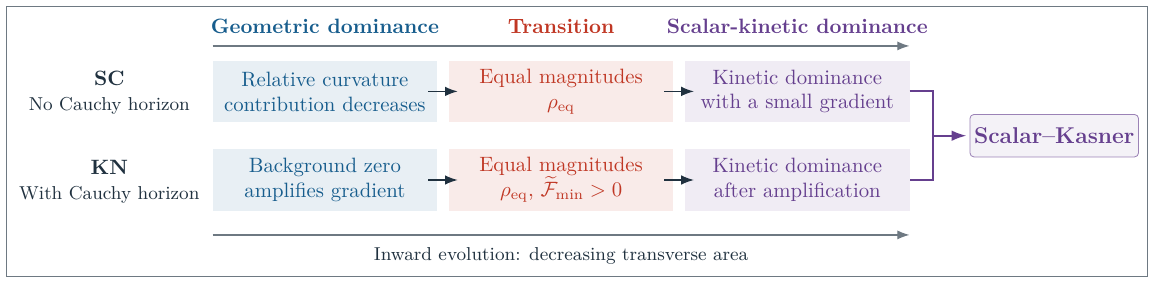}
\caption{Two dynamical routes for small, boundary-supported stationary scalar input as the area of the ingoing null geodesic congruence contracts. On the Schwarzschild-like route (SC), the curvature term decreases relative to the scalar kinetic term while the gradient remains small. The terms have equal magnitude at $\rho_{\rm eq}$ and enter the metric equation with the same sign. On the Kerr--Newman-like route (KN), area contraction toward the would-be Cauchy-horizon zero strongly amplifies the gradient. The transition then transfers dominance to the kinetic term. The opposite-sign terms cancel at $\rho_{\rm eq}$, giving $\widetilde{\mathcal F}_{\min}>0$ in the place of the would-be horizon. Both routes approach scalar--Kasner behavior. Arrows show the sequence, and lengths are schematic.}
\label{fig:stages}
\end{figure*}

We study four-dimensional stationary, axisymmetric interiors with boundary-supported small scalar input. The rotating calculation retains the polar-angle dependence and solves the coupled metric, scalar, and Maxwell fields. A common radial system connects amplification, dominance transfer, and interior geometry along two routes during area contraction (Fig.~\ref{fig:stages}). Without a Cauchy horizon, contraction reduces the curvature contribution relative to the scalar kinetic term. The kinetic term dominates while the gradient is still small, and an order-one departure from the background metric accumulates over a much longer radial interval. With a Cauchy horizon, contraction carries the solution toward the would-be horizon zero, strongly amplifying the gradient before kinetic dominance. The subsequent transition produces a positive minimum of the weighted metric diagnostic, explaining inner-horizon loss dynamically and complementing no-inner-horizon results~\cite{CLY,ALY}. Both routes lead to a local Kasner regime with a scalar field, termed scalar--Kasner below. Fully coupled spherical and stationary axisymmetric calculations test the relative magnitudes of the terms, gradient amplification, positive metric minimum, and late Kasner exponents. The Supplemental Material~\cite{SM} contains the detailed derivations and scale diagram, order estimates and matching analysis, and numerical methods and accuracy checks.

\paragraph{\normalfont\bfseries Model and definitions.}
In four spacetime dimensions, we consider a real electrically neutral scalar $\Psi$ of mass $\mu_s$ minimally coupled to Einstein--Maxwell theory,
\begin{equation}
16\pi\mathcal L=R_4-F_{\mu\nu}F^{\mu\nu}-2(\nabla\Psi)^2-2\mu_s^2\Psi^2 .
\label{lagrangian}
\end{equation}
We set Newton's constant and the speed of light to unity. Here $\mathcal L$ is the Lagrangian density, $R_4$ the Ricci scalar, $F_{\mu\nu}$ the Maxwell field strength, and $\nabla$ the metric covariant derivative. We use the stationary, axisymmetric metric and matter fields
\begin{equation}
\begin{aligned}
ds^2={}&g_{rr}^{\perp}dr^2
+g_{\theta\theta}\left(d\theta+\frac{g_{r\theta}}{g_{\theta\theta}}dr\right)^2\\
&+g_{tt}^{\perp}dt^2+g_{\varphi\varphi}(d\varphi-\Omega\,dt)^2,\\
\Psi={}&\Psi(r,\theta),\qquad A=A_t\,dt+A_\varphi\,d\varphi.
\end{aligned}
\label{metricansatz}
\end{equation}
Here $r,\theta$ are radial and polar coordinates, $t$ is the time coordinate, and $\varphi$ is the azimuthal angle. Stationarity and axisymmetry mean independence of $t$ and $\varphi$, respectively. Completing the two squares defines $g_{rr}^{\perp}=g_{rr}-g_{r\theta}^2/g_{\theta\theta}$, $g_{tt}^{\perp}=g_{tt}-g_{t\varphi}^2/g_{\varphi\varphi}$, and $\Omega=-g_{t\varphi}/g_{\varphi\varphi}$. All coefficients depend only on $(r,\theta)$. $\Omega$ describes frame dragging and $A$ is the electromagnetic potential.

The scalar-free background is Kerr--Newman (KN), with mass $M$, electric charge $Q$, and angular momentum $J$. Its rotation parameter $a=J/M$ is the angular momentum per unit mass. We take a nonextremal background, $M^2>Q^2+a^2$. Reissner--Nordstr\"om (RN) and Kerr are its nonrotating and uncharged limits. Setting $Q=a=0$ gives the Schwarzschild background. We prescribe regular horizon data $\Psi_h=\epsilon\hat\Psi_h$, where $h$ labels the event horizon, $\epsilon$ is the dimensionless signed amplitude, and $\hat\Psi_h$ is a fixed normalized angular profile ($\hat\Psi_h=1$ in spherical symmetry). We call small-amplitude scalar input at the event horizon ($|\epsilon|\ll1$) small scalar input. Scaling with $\epsilon$ is evaluated at fixed background parameters, scalar mass, and normalized profile within this stationary family. To study the deep interior, we instead follow each solution inward at fixed small input.

For the numerical family, a fixed-shape stationary scalar profile at a finite exterior radius supplies horizon-regular linear data for the coupled nonlinear interior.

In a rotating geometry, $r$ need not equal the local areal radius $r_{\rm a}$. To measure contraction, we use an ingoing null geodesic congruence. Let $dA$ be the cross-sectional area of this congruence and $dA_h$ its value at the event horizon. We use the dimensionless coordinate
\begin{equation}
\rho=-\frac12\ln\frac{dA}{dA_h}=-\ln\frac{r_{\rm a}}{r_h}.
\label{rho}
\end{equation}
Here $r_h=\sqrt{r_+^2+a^2}$ is the horizon areal radius, whereas $r_+$ is the background event-horizon value of the radial coordinate. They differ when $a\ne0$. Area contraction corresponds to negative expansion of the ingoing null geodesic congruence. Dots denote derivatives with respect to $\rho$ along a ray with fixed horizon label $\zeta_h=\cos\theta_h$, where $\theta_h$ is its initial polar angle measured from the rotation axis. The ray's physical angle may change as it moves inward.

The would-be Cauchy horizon has position $\rho_i(\zeta_h)$. To track the causal character of surfaces of constant $\rho-\rho_i$, define
\begin{equation}
\mathcal F=-r_h^2\{\nabla[\rho-\rho_i(\zeta_h)]\}^2 .
\label{Fdef}
\end{equation}
Here $(\nabla X)^2=g^{\mu\nu}\partial_\mu X\partial_\nu X$. The gradient is normal to these surfaces. A positive $\mathcal F$ means a timelike normal and a spacelike surface. A null surface requires $\mathcal F=0$. The would-be horizon therefore has $\mathcal F_{\rm ref}=0$, where $\mathrm{ref}$ denotes the scalar-free background. Subtracting $\rho_i(\zeta_h)$ places that horizon at the same zero value on every ray, while retaining its angular dependence. For the SC route, which has no Cauchy horizon, we use $\mathcal F=-r_h^2(\nabla\rho)^2$.

The coefficient $\mathcal C_g$ collects geometric, electromagnetic, scalar-potential, and scalar angular-gradient contributions. Define
\begin{equation}
\widetilde{\mathcal F}=e^{-3\rho}\mathcal F,\qquad
\widetilde{\mathcal C}_g=e^{-3\rho}\mathcal C_g.
\label{weighteddefs}
\end{equation}
We use $\widetilde{\mathcal F}$ as the weighted metric diagnostic. The factor $e^{-3\rho}$ removes the fixed $3\mathcal F$ term from the radial equation and preserves the sign and zeros of $\mathcal F$, but not its extrema. For $\widetilde{\mathcal F}\ne0$, the coupled radial equations take the form
\begin{equation}
\partial_\rho\ln|\widetilde{\mathcal F}|=\dot\Psi^2+\mathcal S_f-\frac{\widetilde{\mathcal C}_g}{\widetilde{\mathcal F}},\quad
\ddot\Psi=\frac{\widetilde{\mathcal C}_g}{\widetilde{\mathcal F}}\dot\Psi+S_\psi .
\label{sourcebalance}
\end{equation}
The positive term $\dot\Psi^2$ is the radial scalar kinetic contribution. The term $\mathcal S_f$ describes focusing due to shear and electromagnetic fields, and $S_\psi$ contains scalar-mass and angular-derivative terms. Geometric dominance here includes all remaining curvature and matter contributions in the metric equation. Angular derivatives in the field equations couple the inward evolution at different polar angles.

To compare these magnitudes, define
\begin{equation}
\mathcal O_\rho=\frac{\dot\Psi^2-|\mathcal S_f|-|\widetilde{\mathcal C}_g/\widetilde{\mathcal F}|}
{\dot\Psi^2+|\mathcal S_f|+|\widetilde{\mathcal C}_g/\widetilde{\mathcal F}|} .
\label{dominance}
\end{equation}

\begin{figure}[t]
\centering\includegraphics[width=\linewidth,trim=0 3bp 0 5bp,clip]{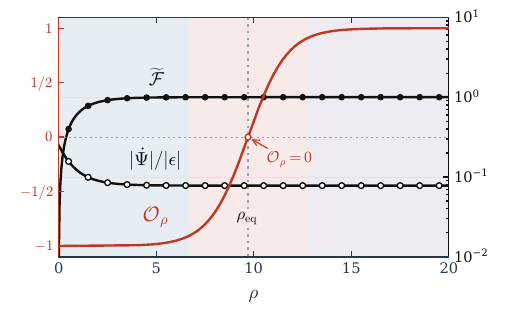}
\caption{Area contraction for $Q=a=0$, $r_h=1$, $\mu_s r_h=0.5$, and $\epsilon=0.1$. Black filled/open circles show $\widetilde{\mathcal F}$ and $|\dot\Psi|/|\epsilon|$ on the logarithmic right axis. Red shows the full $\mathcal O_\rho$ on the left. The dotted line marks $\rho_{\rm eq}\simeq9.7054$. Blue, red, and purple backgrounds divide the display at $\rho_{\rm eq}\pm3$. All curves use the coupled spherical solution.}
\label{fig:neutral}
\end{figure}

Thus $\mathcal O_\rho\to-1$ indicates dominance of the remaining terms and $\mathcal O_\rho\to1$ scalar kinetic dominance. At $\mathcal O_\rho=0$, the kinetic term equals the sum of the other magnitudes. With negligible $\mathcal S_f$, this defines $\rho_{\rm eq}$ in the leading approximation. Equal magnitudes produce cancellation only for opposite signs, as the two routes show below.

\paragraph{\normalfont\bfseries Route I: No Cauchy horizon.}
For spherical horizon data with $Q=a=0$ and $F_{\mu\nu}=0$, the Schwarzschild-like branch has $r_{\rm a}=r$ and $\mathcal S_f=0$. At large $\rho$, the weighted mass terms are exponentially suppressed, and the small-input estimates give $\widetilde{\mathcal C}_g\simeq-e^{-\rho}$ and $S_\psi\simeq0$.
The regular horizon solution gives $\dot\Psi=O(\epsilon)$ at any fixed interior position. Equation~\eqref{sourcebalance} makes the positive $\widetilde{\mathcal F}$ increase, so $e^{-\rho}/\widetilde{\mathcal F}$ decays at least exponentially. Integrating its scalar equation gives
\begin{equation}
\dot\Psi(\rho)\simeq\dot\Psi(\rho_0)
\exp\!\left[-\int_{\rho_0}^{\rho}
\frac{e^{-u}}{\widetilde{\mathcal F}(u)}\,du\right].
\label{neutralgradient}
\end{equation}
Here $\rho_0$ is a fixed interior position in this contraction regime. The integral is finite, so the gradient remains of order $\epsilon$. As the curvature contribution $e^{-\rho}/\widetilde{\mathcal F}$ decays, the scalar kinetic term becomes dominant and $\mathcal O_\rho\to1$.

The scalar kinetic term becomes dominant because the relative curvature contribution decays, even though spacetime curvature grows. At $\rho_{\rm eq}$, dominance passes from curvature to the scalar kinetic term, while $\widetilde{\mathcal F}$ continues to increase (Figs.~\ref{fig:stages} and \ref{fig:neutral}). The equality position is $\rho_{\rm eq}=O(\ln|\epsilon|^{-1})$, with an $O(1)$ transition width. After this transition, the scalar-induced departure from the background accumulates slowly: $\partial_\rho\ln(\widetilde{\mathcal F}/\widetilde{\mathcal F}_{\rm ref})\simeq\dot\Psi^2=O(\epsilon^2)$ once the curvature term has decayed. An order-one change in this logarithmic ratio requires $\Delta\rho=O(\epsilon^{-2})$. The metric itself still contains the rapid geometric factor $\mathcal F=e^{3\rho}\widetilde{\mathcal F}$.

\begin{figure}[t]
\centering\includegraphics[width=\linewidth,trim=0 3bp 0 5bp,clip]{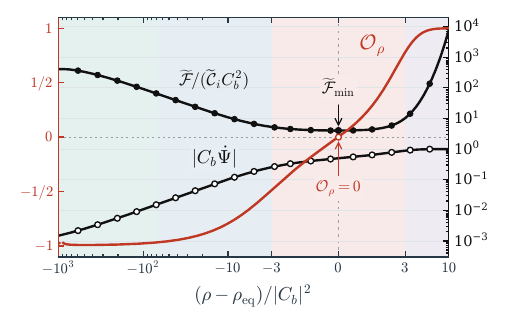}
\caption{KN amplification and transition to kinetic dominance for $Q=0.6/\sqrt2$, $a=0.8/\sqrt2$, $r_+=1$, $\mu_s r_+=0.5$, $\epsilon=0.1$, and $\theta_h=60^\circ$. The right logarithmic axis shows $\widetilde{\mathcal F}/(\widetilde{\mathcal C}_iC_b^2)$ (filled circles) and $|C_b\dot\Psi|$ (open circles). The left axis shows $\mathcal O_\rho$ (red). The horizontal axis shows $(\rho-\rho_{\rm eq})/|C_b|^2$, the distance from $\rho_{\rm eq}$ in units of the transition scale. Dotted guides locate the positive minimum of $\widetilde{\mathcal F}$ and equal magnitudes of the kinetic and remaining terms, $\mathcal O_\rho=0$. The normalized scalar gradient approaches unity, $C_b\dot\Psi\to1$.}
\label{fig:kn}
\end{figure}

\paragraph{\normalfont\bfseries Route II: With Cauchy horizon.}
On the Kerr--Newman-like route, area contraction carries the solution toward the would-be Cauchy horizon. As $\widetilde{\mathcal F}_{\rm ref}$ approaches zero, the scalar gradient is strongly amplified. Near the would-be Cauchy horizon, the linear scalar solution contains a logarithm. Write its coefficient as $C_b=\epsilon b(\zeta_h)$, where the normalized linear solution fixes the finite coefficient $b$. The initial gradient is $O(|C_b|)$, and the near-horizon behavior is
\begin{equation}
\widetilde{\mathcal F}_{\rm ref}\simeq\widetilde{\mathcal C}_i(\rho_i-\rho),
\qquad
\dot\Psi\simeq\frac{C_b}{\rho_i-\rho}.
\label{incoming}
\end{equation}

Here $\widetilde{\mathcal C}_i=-\partial_\rho\widetilde{\mathcal F}_{\rm ref}|_{\rho_i}>0$ is the background slope magnitude. Equation~\eqref{incoming} gives $\Psi\simeq-C_b\ln(\rho_i-\rho)+\mathrm{const.}$ and the incoming flux $\widetilde{\mathcal F}_{\rm ref}\dot\Psi\simeq\widetilde{\mathcal C}_iC_b$.

At $\rho_i-\rho=O(|C_b|)$, the gradient is already $O(1)$ but the kinetic-to-geometric ratio is $O(|C_b|)\ll1$. Amplification therefore precedes kinetic dominance.

Incoming matching gives $\widetilde{\mathcal F}\dot\Psi\simeq\widetilde{\mathcal C}_iC_b$ for the actual fields, hence
\begin{equation}
\frac{\dot\Psi^2}{\widetilde{\mathcal C}_i/\widetilde{\mathcal F}}
\simeq\frac{\widetilde{\mathcal C}_iC_b^2}{\widetilde{\mathcal F}}.
\label{incomingratio}
\end{equation}
The terms become comparable at $\widetilde{\mathcal F}=O(\widetilde{\mathcal C}_iC_b^2)$ and $|\dot\Psi|=O(|C_b|^{-1})$. The incoming evolution therefore establishes the inverse-input gradient scale at the entrance to the layer. Within it, the gradient changes by an order-one factor as dominance passes to the kinetic term. The background expansion supplies matching data, while the actual $\widetilde{\mathcal F}$ evolves through the transition.

On the smooth small-input branch, the layer analysis requires $\widetilde{\mathcal C}_g\simeq\widetilde{\mathcal C}_i$ and subleading normalized focusing and scalar-equation corrections. The scalar terms admit explicit order estimates, while the geometric and Maxwell terms require regularity bounds on the coupled fields. The retained terms in the metric equation are then $O(|C_b|^{-2})$, and those in the scalar equation are $O(|C_b|^{-3})$, selecting the width $O(C_b^2)$. Their common order within each equation allows normalization to remove the input amplitude at fixed $\widetilde{\mathcal C}_i$. This explains the dynamical origin of the emergent scaling symmetry reported for spherical interiors in Ref.~\cite{Sun}, now within equations that retain angular coupling. The leading equations are
\begin{equation}
\partial_\rho\ln|\widetilde{\mathcal F}|
\simeq\dot\Psi^2-\frac{\widetilde{\mathcal C}_i}{\widetilde{\mathcal F}},
\qquad
\ddot\Psi\simeq\frac{\widetilde{\mathcal C}_i}{\widetilde{\mathcal F}}\dot\Psi .
\label{competition}
\end{equation}
The leading equations conserve a combination fixed by the incoming flux in Eq.~\eqref{incoming}:
\begin{equation*}
\dot\Psi+\frac{\widetilde{\mathcal C}_i}{\widetilde{\mathcal F}\dot\Psi}
\simeq C_b^{-1}.
\end{equation*}
Solving for $\widetilde{\mathcal F}$ gives
\begin{equation}
\widetilde{\mathcal F}\simeq
\frac{\widetilde{\mathcal C}_i C_b^2}{(C_b\dot\Psi)(1-C_b\dot\Psi)},
\qquad 0<C_b\dot\Psi<1.
\label{matchedmetric}
\end{equation}
The denominator is positive and has maximum $1/4$ at $C_b\dot\Psi=1/2$. Thus $\widetilde{\mathcal F}$ reaches a positive minimum at $\rho_{\rm eq}$:
\begin{equation}
\begin{gathered}
\left.\partial_\rho\widetilde{\mathcal F}\right|_{\rho_{\rm eq}}=0,
\qquad \dot\Psi_{\rm eq}\simeq\frac{1}{2C_b},\\
\widetilde{\mathcal F}_{\min}
=\widetilde{\mathcal F}(\rho_{\rm eq})
\simeq4\widetilde{\mathcal C}_iC_b^2>0.
\end{gathered}
\label{horizonloss}
\end{equation}
Here $\dot\Psi_{\rm eq}=\dot\Psi(\rho_{\rm eq})$. At the minimum, the positive kinetic term balances the negative geometric term. Geometry drives $\widetilde{\mathcal F}$ downward before equality, and the kinetic term drives it upward afterward. Since $\mathcal F=e^{3\rho}\widetilde{\mathcal F}>0$, surfaces of constant $\rho-\rho_i$ remain spacelike: the amplified gradient replaces the would-be Cauchy-horizon zero with a positive minimum. Figure~\ref{fig:kn} tests the predicted gradient growth, equality, and positive minimum of $\widetilde{\mathcal F}$.

Incoming matching places the minimum at $\rho_{\rm eq}>\rho_i$, with $\rho_{\rm eq}-\rho_i=O(C_b^2\ln|C_b|^{-1})$. Beyond the $O(C_b^2)$ transition layer, the geometric term in the first integral becomes subleading, fixing a stable gradient plateau. Define $\Psi_{\rm out}$ at $\rho_{\rm out}-\rho_{\rm eq}=O(|C_b|)$. The incoming normalization then predicts
\begin{equation}
\dot\Psi_{\rm out}=\dot\Psi(\rho_{\rm out},\zeta_h),\qquad
C_b\dot\Psi_{\rm out}\xrightarrow{\epsilon\to0}1.
\label{gain}
\end{equation}
Thus the radial dynamics fixes the plateau coefficient as well as its inverse-input scale, consistent with spherical results~\cite{vdm,He}. Figure~\ref{fig:kn} numerically tests this prediction, with $C_b\dot\Psi$ approaching unity. Rescaled KN, RN, and Kerr profiles test the common transition shape. Further inward evolution determines the final Kasner exponents.

\paragraph{\normalfont\bfseries A common endpoint: scalar--Kasner.}
The late radial solution gives the Kasner metric once the scalar gradient and local spatial directions settle, with spatial curvature, electromagnetic fields, scalar spatial gradients, and the potential subleading~\cite{LiKasner,Henneaux}. Stationarity and subleading angular derivatives allow the same leading $\rho$ equations to describe evolution along the timelike normal. A polar label $\vartheta$ fixed along this normal removes the radial--polar cross term in Eq.~\eqref{metricansatz}, with $\vartheta\ne\theta$ in general.

In this approximation, Eq.~\eqref{rho} fixes the angular area, $\sqrt{g_{\vartheta\vartheta}g_{\varphi\varphi}}\propto e^{-2\rho}$. The radial trace and the difference of the angular Einstein equations fix the ratio, $g_{\vartheta\vartheta}/g_{\varphi\varphi}\propto e^{-4\gamma\rho}$, with $\gamma$ constant to leading order along the inward evolution. Together they give
\begin{equation}
g_{\vartheta\vartheta}\propto e^{-2(1+\gamma)\rho},\qquad
g_{\varphi\varphi}\propto e^{-2(1-\gamma)\rho}.
\label{transversescales}
\end{equation}
The opposite $\gamma$ terms change the two angular lengths differently without changing the area law. This deformation is shear. Spherical symmetry gives $\gamma=0$. Let $\beta=\lim\dot\Psi$ be the settled scalar gradient. Once electromagnetic focusing is subleading, the shear contribution gives $\mathcal S_f\simeq\gamma^2$. Shear can split the transverse exponents even when $\gamma^2\ll\beta^2$. Equation~\eqref{sourcebalance} gives $\partial_\rho\ln\widetilde{\mathcal F}\simeq\beta^2+\gamma^2$. The radial trace equation and the metric relation also give $g_{tt}^{\perp}\mathcal F\propto e^{4\rho}$. Integrating the first relation and using the second yields
\begin{equation}
\mathcal F\propto e^{(3+\beta^2+\gamma^2)\rho},\qquad
g_{tt}^{\perp}\propto e^{(1-\beta^2-\gamma^2)\rho}.
\label{lateradialmetric}
\end{equation}
Along the timelike normal, the line element is $ds^2\simeq-r_h^2\,d\rho^2/\mathcal F$. Thus $\mathcal F$ converts area contraction into proper-time evolution. The remaining proper time $\tau$ decreases as $\rho$ increases:
\begin{equation}
\frac{d\tau}{d\rho}\simeq-\frac{r_h}{\sqrt{\mathcal F}},\qquad
\tau\propto e^{-(3+\beta^2+\gamma^2)\rho/2}.
\label{radialtopropertime}
\end{equation}
Thus the exponentials in $\rho$ become powers of $\tau$. With the limiting angular shift $\Omega_*=\lim\Omega$, the metric and scalar take the form
\begin{equation}
\begin{aligned}
ds^2\simeq{}&-d\tau^2+C_t^2(\tau/\tau_0)^{2p_t}dt^2\\
&+C_\theta^2(\tau/\tau_0)^{2p_\theta}d\vartheta^2\\
&+C_\varphi^2(\tau/\tau_0)^{2p_\varphi}(d\varphi-\Omega_*dt)^2,\\
\Psi\simeq{}&\Psi_0+p_\Psi\ln(\tau/\tau_0).
\end{aligned}
\label{kasner}
\end{equation}
Here $t$ labels a spatial direction and $\tau\to0$ at the singularity. The constants $\tau_0>0$, $C_j>0$, and $\Psi_0$ fix reference scales and the scalar offset. The exponents are
\begin{equation}
\begin{aligned}
p_t&=\frac{\beta^2+\gamma^2-1}{3+\beta^2+\gamma^2},\qquad
p_\Psi=-\frac{2\beta}{3+\beta^2+\gamma^2},\\
p_\theta&=\frac{2(1+\gamma)}{3+\beta^2+\gamma^2},\qquad
p_\varphi=\frac{2(1-\gamma)}{3+\beta^2+\gamma^2}.
\end{aligned}
\label{kasnerexponents}
\end{equation}
They satisfy $\sum_jp_j=1$ and $\sum_jp_j^2+2p_\Psi^2=1$, with $j=t,\theta,\varphi$ and the factor 2 fixed by Eq.~\eqref{lagrangian}. Positive spatial exponents mean contraction as $\tau\to0$. On the SC route, $\gamma=0$ and $\beta=O(\epsilon)$, so for small scalar input the exponents are close to $(-1/3,2/3,2/3;0)$. The KN plateau supplies a gradient of order $|C_b|^{-1}$. A bounded accumulated correction in the subsequent scalar evolution preserves this order in the final $\beta$. If shear remains subleading to the scalar kinetic term, $\gamma^2/\beta^2\to0$ for small input, giving the late $\mathcal O_\rho\to1$. The continued solutions test these conditions, exponent settling, and suppression of the neglected terms. Independent radial-slope estimates of $\beta,\gamma$ reproduce the four Kasner exponents.

\paragraph{\normalfont\bfseries Summary and outlook.}
For small, boundary-supported stationary scalar input, area contraction drives a common transition from geometric to scalar kinetic dominance. Without a Cauchy horizon, the scalar kinetic term becomes dominant while the gradient remains small, long before a large metric response accumulates. With a Cauchy horizon, contraction toward the metric zero establishes a large gradient before a narrow layer completes the change in dominance. On this route, cancellation between the kinetic and geometric terms produces a positive minimum of $\widetilde{\mathcal F}$ and explains inner-horizon loss dynamically. Both routes lead to scalar--Kasner interiors. For finite or time-dependent input, it remains to determine whether the same transition scales and KN amplification law apply. Other matter fields, more general couplings, and higher derivatives also raise the question of whether the interior still approaches a scalar--Kasner regime.

\textbf{Data availability.---} Source files and two standalone numerical programs accompany this Letter and its Supplemental Material.

\paragraph{\normalfont\bfseries Acknowledgments.}
The author thanks Li Li, Run-Qiu Yang, and Ze Sun for helpful discussions.


\begin{thebibliography}{99}
\bibitem{PI} E. Poisson and W. Israel, Internal structure of black holes, Phys. Rev. D \textbf{41}, 1796 (1990), \href{https://doi.org/10.1103/PhysRevD.41.1796}{\nolinkurl{doi:10.1103/PhysRevD.41.1796}}.
\bibitem{Ori} A. Ori, Inner structure of a charged black hole: An exact mass-inflation solution, Phys. Rev. Lett. \textbf{67}, 789 (1991), \href{https://doi.org/10.1103/PhysRevLett.67.789}{\nolinkurl{doi:10.1103/PhysRevLett.67.789}}.
\bibitem{Dafermos} M. Dafermos, The interior of charged black holes and the problem of uniqueness in general relativity, Commun. Pure Appl. Math. \textbf{58}, 445 (2005), \href{https://arxiv.org/abs/gr-qc/0307013}{\nolinkurl{arXiv:gr-qc/0307013}}.
\bibitem{HA} A. J. S. Hamilton and P. P. Avelino, The physics of the relativistic counter-streaming instability that drives mass inflation inside black holes, Phys. Rep. \textbf{495}, 1 (2010), \href{https://arxiv.org/abs/0811.1926}{\nolinkurl{arXiv:0811.1926}}.
\bibitem{DL} M. Dafermos and J. Luk, The interior of dynamical vacuum black holes I: The $C^0$-stability of the Kerr Cauchy horizon, Ann. Math. \textbf{202}, 309 (2025), \href{https://arxiv.org/abs/1710.01722}{\nolinkurl{arXiv:1710.01722}}.
\bibitem{Gurriaran} S. Gurriaran, Non-linear instability of the Kerr Cauchy horizon near $i_+$, \href{https://arxiv.org/abs/2603.17911}{\nolinkurl{arXiv:2603.17911}} (2026).
\bibitem{GuoJoshi} J.-Q. Guo and P. S. Joshi, Interior dynamics of neutral and charged black holes, Phys. Rev. D \textbf{92}, 064013 (2015), \href{https://arxiv.org/abs/1507.01806}{\nolinkurl{arXiv:1507.01806}}.
\bibitem{Frenkel} A. Frenkel, S. A. Hartnoll, J. Kruthoff, and Z. D. Shi, Holographic flows from CFT to the Kasner universe, J. High Energy Phys. \textbf{08} (2020) 003, \href{https://arxiv.org/abs/2004.01192}{\nolinkurl{arXiv:2004.01192}}.
\bibitem{CLY} R.-G. Cai, L. Li, and R.-Q. Yang, No inner-horizon theorem for black holes with charged scalar hairs, J. High Energy Phys. \textbf{03} (2021) 263, \href{https://arxiv.org/abs/2009.05520}{\nolinkurl{arXiv:2009.05520}}.
\bibitem{ALY} Y.-S. An, L. Li, and F.-G. Yang, No Cauchy horizon theorem for nonlinear electrodynamics black holes with charged scalar hairs, Phys. Rev. D \textbf{104}, 024040 (2021), \href{https://arxiv.org/abs/2106.01069}{\nolinkurl{arXiv:2106.01069}}.
\bibitem{DPEMH} D. O. Devecio\u glu and M.-I. Park, No scalar-haired Cauchy horizon theorem in Einstein--Maxwell--Horndeski theories, Phys. Lett. B \textbf{829}, 137107 (2022), \href{https://arxiv.org/abs/2101.10116}{\nolinkurl{arXiv:2101.10116}}.
\bibitem{DPEMGB} D. O. Devecio\u glu and M.-I. Park, No scalar-haired Cauchy horizon theorem in charged Gauss--Bonnet black holes, Eur. Phys. J. C \textbf{84}, 168 (2024), \href{https://arxiv.org/abs/2307.10532}{\nolinkurl{arXiv:2307.10532}}.
\bibitem{Hartnoll} S. A. Hartnoll, G. T. Horowitz, J. Kruthoff, and J. E. Santos, Diving into a holographic superconductor, SciPost Phys. \textbf{10}, 009 (2021), \href{https://arxiv.org/abs/2008.12786}{\nolinkurl{arXiv:2008.12786}}.
\bibitem{DHS} O. J. C. Dias, G. T. Horowitz, and J. E. Santos, Inside an asymptotically flat hairy black hole, J. High Energy Phys. \textbf{12} (2021) 179, \href{https://arxiv.org/abs/2110.06225}{\nolinkurl{arXiv:2110.06225}}.
\bibitem{li} W. Li and M. Van de Moortel, Kasner bounces and fluctuating collapse inside hairy black holes with charged matter, Ann. PDE \textbf{11}, 3 (2025), \href{https://arxiv.org/abs/2302.00046}{\nolinkurl{arXiv:2302.00046}}.
\bibitem{LSY} L. Li, Z. Sun, and F.-G. Yang, Critical phenomenon inside asymptotically flat black holes with spontaneous scalarization, Phys. Rev. Lett. \textbf{136}, 251402 (2026), \href{https://arxiv.org/abs/2512.19377}{\nolinkurl{arXiv:2512.19377}}.
\bibitem{Gao} L.-L. Gao, Y. Liu, and H.-D. Lyu, Critical Kasner scaling from Cauchy-horizon collapse near holographic phase transitions, \href{https://arxiv.org/abs/2609.24227}{\nolinkurl{arXiv:2609.24227}} (2026).
\bibitem{BHR} Y. Brihaye, C. Herdeiro, and E. Radu, Inside black holes with synchronized hair, Phys. Lett. B \textbf{760}, 279 (2016), \href{https://arxiv.org/abs/1605.08901}{\nolinkurl{arXiv:1605.08901}}.
\bibitem{GLL} L.-L. Gao, Y. Liu, and H.-D. Lyu, Internal structure of hairy rotating black holes in three dimensions, J. High Energy Phys. \textbf{01} (2024) 063, \href{https://arxiv.org/abs/2310.15781}{\nolinkurl{arXiv:2310.15781}}.
\bibitem{vdm} M. Van de Moortel, Violent nonlinear collapse in the interior of charged hairy black holes, Arch. Ration. Mech. Anal. \textbf{248}, 89 (2024), \href{https://arxiv.org/abs/2109.10932}{\nolinkurl{arXiv:2109.10932}}.
\bibitem{Henneaux} M. Henneaux, The final Kasner regime inside black holes with scalar or vector hair, J. High Energy Phys. \textbf{03} (2022) 062, \href{https://arxiv.org/abs/2202.04155}{arXiv:2202.04155}.
\bibitem{SM} See Supplemental Material for the exact radial equations, the two small-scalar-input routes, scalar--Kasner asymptotics, numerical results, computational equations, and accuracy checks. It also includes Refs.~\cite{MW,AnderssonRendall,GJ}.
\bibitem{He} C. He and C. Liu, Kasner critical scaling in Einstein--Maxwell--scalar black holes with nonlinear collapse and scalarization, \href{https://arxiv.org/abs/2609.35500}{\nolinkurl{arXiv:2609.35500}} (2026).
\bibitem{Sun} Z. Sun, F.-G. Yang, R.-G. Cai, L. Li, and R.-Q. Yang, Emergent scaling symmetry and universal interior criticality of scalarized black holes, \href{https://arxiv.org/abs/2609.30240}{\nolinkurl{arXiv:2609.30240}} (2026).
\bibitem{LiKasner} W. Li, Kasner-like description of spacelike singularities in spherically symmetric spacetimes with scalar matter, Ann. PDE \textbf{11}, 5 (2025), \href{https://arxiv.org/abs/2304.04802}{\nolinkurl{arXiv:2304.04802}}.
\bibitem{MW} T. M\"adler and J. Winicour, Bondi--Sachs formalism, Scholarpedia \textbf{11}, 33528 (2016), \href{https://arxiv.org/abs/1609.01731}{\nolinkurl{arXiv:1609.01731}}.
\bibitem{AnderssonRendall} L. Andersson and A. D. Rendall, Quiescent cosmological singularities, Commun. Math. Phys. \textbf{218}, 479 (2001), \href{https://arxiv.org/abs/gr-qc/0001047}{\nolinkurl{arXiv:gr-qc/0001047}}.
\bibitem{GJ} E. Gourgoulhon and J. L. Jaramillo, A $3+1$ perspective on null hypersurfaces and isolated horizons, Phys. Rep. \textbf{423}, 159 (2006), \href{https://arxiv.org/abs/gr-qc/0503113}{arXiv:gr-qc/0503113}.
\end{thebibliography}
\end{document}